# SECURITY PROTOCOLS FOR DISTRIBUTED WIRELESS SENSOR NETWORKS

## Ovidiu Baniaş, Daniel-Ioan Curiac

Automatics and Applied Informatics Department, "Politehnica University of Timisoara", Automation and Computers Faculty, Vasile Parvan Nr. 2, 12345 Timisoara, Romania
Phone: (+40) 256-40-3251, Fax: (+40)-256-40-3214, E-Mail: ovidiu.banias@aut.upt.ro,

** Department of Significance, Major Company in Industry, Inc., Postal address, 56789 OtherCity, OtherCountry
Phone: (567) 999-9999, Fax: (567) 999-1111, E-Mail: e.co-author@major.com, WWW: http://www.major.com/~coauthor

*Abstract* – *Sensor Networks technologies had proved their great practicability in the real world, being just a matter of time until this kind of networks will be standardized and used in the field. We focus on security issues in Distributed Sensor Networks like useful cryptosystems, attacks, preventing and detecting possible attacks, describing state of the art in domain and proposing new methods for further investigation.*

## I. INTRODUCTION

Wireless sensor networks are one of the most important and in development technologies of the 21 century, being early researched for military purposes in 1980s, 1990s. Recent growth of technology (increased computational power, smaller chips, microprocessors with less power consumption) opened a new world for research in this domain. Also, more research groups are activating today: Defense Advanced Research Projects Agency (DARPA) with two important projects: Distributed Sensor Networks (DSN) and Sensor Information Technology (SensIT) [1], Berkeley WEBS, Crossbow, etc.

A sensor network is a network of small distributed devices that are using sensors for measurement (temperature, motion, pressure, sound) and for prediction (weather forecast, fire ignition, earthquakes, military attack, building safety). These small devices are called motes and are composed of board, sensors, radio transmitter/receptor and power supply (batteries in most cases), some of them being equipped with a small processor for data processing and memory for storage. The motes are deployed in different kind of environments being able to self organize themselves in a hierarchical sensor network and produce the gathered information to the receiver. The main advantage is the small size of the motes (the size of a coin) combined with the small cost per piece and the capacity to self organize not depending of a previous setup and configuration.

The main units of the mote's architecture described in Fig.1 are: power supply unit, sensing unit, processing unit and radio transceiver unit. Power supply unit is composed mostly from disposable batteries or scavenging solar cells. The sensing unit receives the information from the environment through the sensors and then with the ADC (Analog to Digital Converter) transformed sensed information in a digital format to pass forward to the processing unit. The processing unit is the brain of the mote, being composed from a micro processor running on a real time operating system, memory and algorithms. Sensed information is analyzed and processed, possibly stored in the memory, and then according to established protocols, sent through the radio transmitter to neighbor motes.

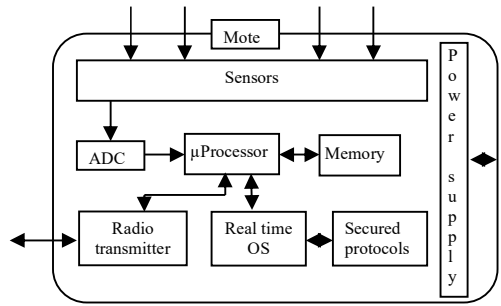

Fig. 1. Mote's architecture

Because motes are running on batteries, the life of the motes and of the sensor network strictly depends of the energy saving design and implementation of communication protocols. Minimizing as low as possible the used resources, raise the problem of security efficiency, the more energy savings the less security for the sensor network. Also, transmitting a bit from a mote to another consumes about the same power needed for executing 800-1000 instructions [2]. So, due to these problems, depending of the domain of interest, amount of information transmitted through the network, degree of needed security, the design of sensor networks should compromise between power consumption and security, depending of specific requirements.

## II. NEED FOR SECURITY IN WIRELESS SENSOR NETWORKS

The design of wireless sensor networks faces two very important problems nowadays: restrictive power consumption and securing communication protocols. Even more protocols had been proposed by researchers, none can be used as a standard and none can guaranty a very good security.

The motes being deployed in a certain environment without any previous network setup should be able to communicate secure and be able to avoid intruders. For this reason cryptographic algorithms should be helpful, but unfortunately not enough for the moment. The sensor network designer should decide on more security implying high power consumption and less security with low power consumption. For example, in wireless sensor networks used for detection (fire, flood, object movement, etc.) the information validity should be no more than few seconds, enough to be propagated to the sink (the sink of a wireless sensor network is a device that reports gathered information to monitoring groups or fire an alarm). In this situation, on one hand, the level of information security is not very high, and could be used symmetric cryptographic algorithms or asymmetric algorithms with shorter key length, but on the other hand, authentication should be very important in order to minimize the possibility an attacker to send wrong information.

Depending on the importance of the transmitted information through the motes, the transmission splits into: reliable and unreliable [3]. Unreliable transmission doesn't care if the information sent to the next mote arrives unaltered or not, because the information was not relevant for the scope of the sensor network (i.e. in a sensor network detecting fire ignition, information "no fire" won't be important because is the default status of the network). On the other hand, important information that change the status of the sensor network and that could cause an alarm should be very careful secured. Also, the network should be able to recognize intruders and not use their information.

The need for securing communication protocols is evident because we can obtain novel solutions for sensor networks self organization and aggregation, but without a high level of security, this kind of networks could not exist and no one would risk money on an easy to break technology.

III SECURING PROTOCOLS AND ATTACKS FOR SENSOR NETWORKS

There are two types of motes architecture: for general use and for dedicated use. The general use motes are meant to be reprogrammed every time is needed (to be used in another sensor network for example). This type of architecture is the nowadays tendency, but leave a wide door opened for attacks. Why? Because being reprogrammable, once an attacker captures one of these motes, will be able to read protocols, keys, and everything, and also will be able to write new software for malicious purposes. On the other hand, the architecture of the dedicated motes is more robust and also more expensive, the protocols and secret keys are all secured in the embedded system (mote), without any possibility for further software reading & rewriting and in case of detecting one of these actions the mote will autodestruct (tamper proof hardware)[4]. In the case of general use motes, once a mote has been captured, the sensor network won't be able by any means to discover the mote is used or not for listening or worse for spoofing. For this reason depending of the deploying environment, should be taken an important decision if using rewritable motes or not.

In [4], it is demonstrated how the EEPROM, program Flash and the SDRAM of a captured mote could be read in less than 1 minute and the security key extracted easily from the SDRAM. Due to this observation, if an attacker could read the mote's hardware is just a matter of time until will find the security key, so we cannot say our data is secured as long as a mote could be captured and read information. The possibility of capturing will never disappear, so the only chance to secure data is to avoid extracting it from the mote, or at least try to postpone the time needed by an attacker to find the security keys. Some work was done in this direction [5][6], proposing an architecture where the motes are deployed with predistributed keys, and later erase those keys and replace with others, and so on.

Security Protocols for Sensor Networks (SPINS), has two secure blocks: SNEP and µTesla. SNEP provides: data confidentiality, two-party data authentication, and data freshness; µTesla is a protocol witch provides authenticated broadcast. Sensor Network Encryption Protocol (SNEP) provides more advantages like semantic security (asymmetric cryptographic algorithms), data authentication and freshness with only 8 bytes added to the message [7]. For authentication the protocol uses a shared counter between sender and receiver (this way there is no need to transmit this peace of information with the message, the counter being incremented with each message) and a MAC (message authentication code). µTesla broadcast protocol it is based also on asymmetric method, with the novel idea of obtaining asymmetry through delaying the disclosure of symmetric keys. To send an authenticated packet the base station computes the MAC of the packet with a key secret key at the moment and than sends the packet to the mote. After the mote receive the packet, it can verify if the secrete key was not disclosed yet, meaning the packet is secured and can be stored in the buffer. At the time the base station publishes the secret key, the MAC of the message stored in the mote's buffer is computed and verified for authenticity [7].

Ad hoc On-demand Distance Vector routing protocol (AODV) and Dynamic Source Routing protocol (DSR), being under research by IETF (Internet Engineering Task Force) for standardization, are good for network traffic but not secured enough (see Table 1) [10].

| Attack | DSR | AODV |
|---|---|---|
| spoofing | yes | yes |
| modification of sequence numbers | no | yes |
| modification of hop counts | no | yes |
| modification of source routes | yes | no |
| tunneling | yes | yes |

Table 1. Attacks on DSR and AODV protocols

TinySec is a communication protocol developed by UC Berkeley research team, is addressed to sensor networks architecture with extreme resource constraints. It uses semantic secure encryption and supports two options: authenticated encryption (TinySec-AE) and authentication only (TinySec-Auth) [11]. But neither this protocol is secure enough, being weak on bogus routing information, selecting forwarding, sinkholes, Sybil, wormholes, and Hello floods [12].

The communication protocols security for sensor networks should be obtain by combining asymmetric and symmetric cryptography with novel design of protocols. The simple use of symmetric or asymmetric cryptography is not enough nowadays. Symmetric key cryptography (Rijandel, RC6, Serpent,..) is fast, but the attacker can forge the secret key quite easy by capturing one of the motes and then using a laptop computer for debugging or brute force attack. On the other hand, asymmetric cryptography (RSA, ECC algorithms) offer a greater security, but is not fast enough for current needs of sensor networks. The National Institute of Standards and Technology (NIST), has proposed Rijandel symmetric algorithm to be the Advanced Encryption Standard(AES), also being the most indicated for 8-bit processors working on the mote technology today. Rijandel outruned TwoFish, RC6, Serpent and MARS, being much faster [13] and better to hardware implement on embedded systems (see Table).

| Symmetric cryptographic algorithm | Clocks | Normalized value |
|---|---|---|
| MARS | 34.163 | 0.28 |
| RC6 | 32.731 | 0.29 |
| RIJANDEL | 9.464 | 1.00 |
| SERPENT | 126.074 | 0.08 |
| TWOFISH | 26.500 | 0.36 |

Table 2. Comparison of symmetric cryptographic algorithms by NIST

Asymmetric algorithms like RSA and ECC are used nowadays for securing systems with good computation power, unfortunately we can not say the same thing about sensor networks where computation power is reduced to minimum for energy saving. Even based on same principles (NP problems), ECC is much better than RSA for small devices like motes of sensor networks [14]. The key length of current RSA offered security is 1024 bits, while same security is offered by ECC at only 160 bits and in this environment, any bit being very important we can state that ECC is the way to go for energy savings (see Table 3). To have a solid security for sensor networks, except using tamper proof hardware, asymmetric algorithm must be used because they can offer authenticity and confidentiality.

Sensor networks due to their restrictive constraints are vulnerable to a lot of attacks like: spoofing, selective forwarding, sinkhole attack, wormhole attack, Sybil attack, Hello flood attack. An attacker could be able to spoof, replay or alter the intercepted information and could create routing loops and generate false messages. Through selective forwarding an intruder can selectively drop certain packages, forwarding just a part of them. This attack could be transformed in a black hole attack if all packages are dropped. The most difficult to detect attack is the combination between sinkhole and wormhole attacks. Sinkhole attack tries to attract packages from neighbor motes with the scope of dropping them and pass forward the information received. In wormhole attack, two distant malicious nodes linked by a low latency band that is not recognized by the sensor network, are placed one near the base station and other near the target, creating a sinkhole and making nodes at a distance from each other to believe they are close. This attack is very powerful because the attacker doesn't need to capture any node from the sensor network, just to deploy two malicious nodes, one near the sink and one near the target. Two schemes were proposed for detecting this kind of attacks but no one proved feasible enough [15]. The Sybil attack [16], a malicious node gathers identities of several nodes in the scope of changing its own identity with the identity of that group of nodes. These identities are later used to break cryptosystems with shared secret keys between multiple nodes. The Hello flood attack sends a lot of "Hello" messages to as many nodes as possible with the scope of shortening battery power.

| Asymmetric cryptographic algorithm | Signature | | Key Exchange | |
|---|---|---|---|---|
| | Sign | Verify | Client | Server |
| RSA-1024 | 304 | 11.9 | 15.4 | 304 |
| ECC-160 | 22.82 | 45.09 | 22.3 | 22.3 |
| RSA-2048 | 2302.7 | 53.7 | 57.2 | 2302.7 |
| ECC-224 | 61.54 | 121.98 | 60.4 | 60.4 |

Table 3. Energy cost in mJ of RSA and ECC algorithms

IV. DETECTING ATTACKS IN SENSOR NETWORKS

Detecting anomalies and intruders in sensor networks is very important. After detection, the sensor network can take decisions to investigate, find and remove malicious nodes if possible. Attacks that only listen to messages are not a big problems for a certain type of sensor networks (detection, alarm reporting) because of the message short time of validity. No attacker can use information like: fire ignition signal, leaking of chemical liquids, traffic problems, building problems, environment monitoring, etc., except for knowing before somebody else finds out or before it is made public (in most cases not being confidential). The most important security problems of sensor networks are spoofing (attacker masquerades as a node in the network, falsifies received messages and sends forward falsified message) and denial of services (DOS).

In the literature there are few proposals about avoiding DOS attacks, an interesting point of view is that neighbor nodes to a jammed area should collaborate and make a border around it and route packages in network according to this border [17]. Detecting node capturing could be done easily by attaching a GPS receiver and in the moment the coordinates changes, the gathered information should be erased automatically and depend the network needs maybe the security keys too. Unfortunately GPS receivers are to expensive to integrate intro sensor networks node, making the overall costs to be too high.

A good way of detecting spoofing attack is the base station from time to time to send packages through the sensor network and after a certain time to verify if message was not altered (Fig. 2). The checking package sent by the base station/sink, will be propagated through the network. If there is a spoofing attacker within the network, the package will be forged and resent. After a certain period of time, the base station will be able to verify if it received any altered package and of course if the network is in danger for attacks.

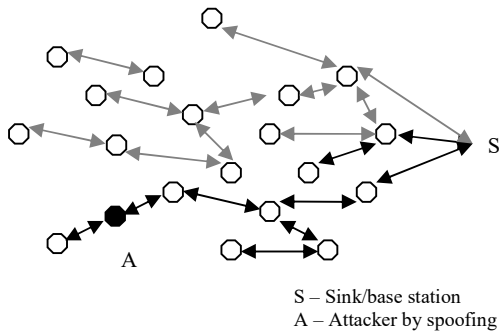

S – Sink/base station
A – Attacker by spoofing

Fig. 2. Detecting spoofing attack

This kind of methods for detecting intruders are important to be researched and implemented because even if will have very good secured protocols (that doesn't exist at the moment) it is always recommended to check from possible attacks from time to time.

## V. IMPROVMENTS

Future work in security for sensor networks, can be done on: researching proper cryptographic algorithms according to desired security level, preventing and detecting attacks, novel communication protocol addressed to different kind of environments and security levels. The trend nowadays is to develop general self organizing sensor networks, witch leaves the possibility to rewrite software on the motes. For this reason it is hard to define security levels and the biggest problem is the possibility an attacker to capture a mote, read and then rewrite the software for malicious purposes. Tamper proof hardware is too expensive, but could be researched software technologies to erase information in case of attack or even to autodestruct themselves. Also Elliptic Curve Criptography (ECC) technologies could become a standard in security for restrictive systems like sensor networks, being the most promising asymmetric algorithm nowadays, unfortunately for the moment it doesn't offer a small enough secret key length.

## VI. CONCLUSION

Researching security for sensor networks should be at least as important as self organization, aggregation and communication protocols. Without security the sensor networks can not be trusted and won't have practical implementation in real world. Researching symmetric and asymmetric cryptographic algorithms, and also preventing and detecting possible attacks should be the next step.